# Speed dependent stochasticity capacitates Newell model for synchronized flow and oscillation growth pattern

Junfang Tian[1], Rui Jiang[2*], Bin Jia[2*], Shoufeng Ma[1], Ziyou Gao[2]

[1]*Institute of Systems Engineering, College of Management and Economics, Tianjin University, Tianjin 300072, China*

[2]*MOE Key Laboratory for Urban Transportation Complex Systems Theory and Technology, Beijing Jiaotong University, Beijing 100044, China*

This paper has incorporated the stochasticity into the Newell car following model. Three stochastic driving factors have been considered: (i) Driver's acceleration is bounded. (ii) Driver's deceleration includes stochastic component, which is depicted by a deceleration with the randomization probability that is assumed to increase with the speed. (iii) Vehicles in the jam state have a larger randomization probability. Two simulation scenarios are conducted to test the model. In the first scenario, traffic flow on a circular road is investigated. In the second scenario, empirical traffic flow patterns in the NGSIM data induced by a rubberneck bottleneck is studied, and the simulated traffic oscillations and synchronized traffic flow are consistent with the empirical patterns. Moreover, two experiments of model calibration and validation are conducted. The first is to calibrate and validate using experimental data, which illustrates that the concave growth pattern has been quantitatively simulated. The second is to calibrate and cross validate vehicles' trajectories using NGSIM data, which exhibits that the car following behaviors of single vehicles can be well described. Therefore, our study highlights the importance of speed dependent stochasticity in traffic flow modeling, which cannot be ignored as in most car-following studies.

## 1. Introduction

Nowadays, traffic congestion is one of the most severe problems that undermine the operation efficiency of modern cities. To reveal the formation and evolution features of traffic congestion in uninterrupted flow, many field observations and measurements have been conducted and various complex spatiotemporal phenomena, such as capacity drop, phantom jam, synchronized traffic flow, have been observed (Chowdhury, 2000; Helbing, 2001; Nagatani, 2002; Kerner, 1998, 2004, 2009; Schönhof and Helbing,2007; Treiber and Kesting, 2013; Jin et al., 2015; Zheng, 2014; Yuan et al., 2017; Arnesen and Hjelkrem, 2018). In order to explain these phenomena, a huge number of traffic flow models were proposed, which have been classified by Kerner into the two-phase models (Helbing, 2001; Nagel et al., 2003; Ni et al. 2016; Pariota et al. 2016; Treiber and Kesting, 2013) and the three-phase models (Kerner, 2013 and 2016).

Two-phase models usually presume that there is a unique relationship between the flow rate and traffic density under the steady state condition. For example, in the Optimal Velocity Model (OVM, Bando et al. 1995) and the Lighthill-Whitham-Richards model (Lighthill and Whitham, 1955; Richards, 1956), the optimal velocity function and the equilibrium speed density relationship explicitly depict this unique relationship, respectively. In Two-phase models, traffic flow is classified into the free flow phase and congested flow phase. Phase transition involved corresponds to a transition from free flow to jams.

---

* Corresponding author
Email address: jftian@tju.edu.cn, jiangrui@bjtu.edu.cn, bjia@bjtu.edu.cn

Based on a long-term empirical data analysis, Kerner introduced the Three-Phase Theory (KTPT, Kerner, 1998, 2004, 2009), which classifies the congested traffic into the synchronized flow (S) and the wide moving jams (J). In congested traffic flow, when speed is not small (or the density is not large), jam will not appear spontaneously. This "synchronized flow" is stable. With decrease of speed (or the increase of density), synchronized flow becomes unstable and jam will appear spontaneously. Phase transitions involved usually correspond to that from free flow (F) to synchronized flow and that from synchronized flow to wide moving jams, which are essentially different from Two-phase theory. As a result, the transition from free flow to jams usually corresponds to a process of F→S→J. In order to reproduce synchronized traffic flow, a variety of complex models have been proposed, most of which are cellular automaton models (Kerner et al., 2002, 2011; Lee et al., 2004; Jiang and Wu, 2003, 2005; Tian et al., 2009, 2014, 2016; Gao et al., 2007, 2009).

Jiang et al. (2014, 2015) carried out a car following experiment on a 3.2-km-long open road section, in which a platoon of 25 passenger cars has been studied. The leading vehicle was asked to move with constant speed. The formation and development of traffic oscillations have been observed. The results showed that the standard deviations of speed increase in a concave or linear way along the 25-car-platoon. For the latter case, due to the physical limits of speeds, unconditional concavity, i.e., a decreasing increment of the amplitude from car to car is expected for sufficiently large platoons. Later, the concave growth pattern of oscillations has been verified by the empirical NGSIM data (Tian, et al., 2016). Moreover, Jiang et al. (2014, 2015) have shown that (i) the simulation results of the Two-phase models, such as General Motor Models (GMs, Gazis et al., 1961), Gipps' Model (Gipps, 1981), OVM (Bando et al., 1995), Full Velocity Difference Model (FVDM, Jiang et al., 2001) and Intelligent Driver Model (IDM, Treiber et al., 2000), run against this finding since the standard deviation initially increases in a convex way in the unstable density range; (ii) by removing the fundamental assumption of unique relationship in steady state in two-phase models and allowing the traffic state to span a two-dimensional region in velocity-spacing plane, the growth pattern of disturbances has changed and becomes qualitatively in accordance with the observations.

The importance of the concave growth of traffic oscillations lies in that it indicates that the instability mechanism of traditional two-phase models is debatable. In traditional models, the unstable traffic flow is generated due to linear instability of the steady state solution. As proved by Li et al. (2014), the linear instability in a nonlinear Newell model leads to initial convex growth of oscillations. We further demonstrate that the linear instability in general two-phase models leads to initial convex growth of oscillations (Tian et al., 2016). Thus, the initial convex growth pattern in Two-phase models contradicts with the observed concave growth pattern, which implies that the mechanism triggering traffic jams in two-phase models is questionable.

Recently, KTPT has been generalized by Tian et al. (2017), in which it is proposed that a model is potentially able to reproduce the evolution of traffic flow, provided the traffic state can dynamically span a 2D region in the speed-spacing plane, no matter there exists steady state or not, or the steady state occupies a 2D region, or a unique relationship exists between speed and spacing. Newell model (Newell, 2002) is a simple but well-known car following model, which assumes that the trajectory of the following vehicle is the same as the leading vehicle, with a time delay and space displacement. Note that this model is different from another one proposed by Newell in 1961 (Newell, 1961). However, Newell model is stable and not able to span a 2D region, and therefore fails to depict growth characteristics of traffic oscillations. Although several improved models (Laval and Leclercq, 2010; Chen et al., 2012a, b; Laval et al., 2014; Chen et al. 2014a, b) have been proposed and claimed to be able to reproduce the formation and propagation of stop-and-go waves, these improved models fail to reproduce other aspects of traffic flow, say, e.g., synchronized flow.

This paper introduces the speed dependent stochasticity into the Newell model. It is shown that speed dependent stochasticity enables the Newell model to reproduce the synchronized flow and concave growth pattern as well. Our study thus highlights the important role of stochasticity in traffic flow modeling. Unfortunately, this factor is ignored in most car-following studies. The paper is organized as follows. Section 2 conducts a brief review of the Newell car following model and its improved versions. Section 3 presents the new model. Simulations results are illustrated in sections 4 and 5. Section 6 gives the calibration and validation results to test performance of the new model. Conclusions are given in Section 7.

## 2. Review of Newell car following model and improved versions

Newell car following model (abbreviated as NCM, see Newell, 2002) assumes that if the vehicles move on a homogeneous road segment, the trajectory of the following vehicle is the same as the leading vehicle, with a translation in time and space, see Figure 1 (a). It is supposed that the vehicle $n$-1 is ahead of vehicle $n$, and their location and speed are $x_{n-1}$ and $x_n$, $v_{n-1}$ and $v_n$, respectively. When vehicle $n$-1 changes its speed from $v$ to $v'$, vehicle $n$ will adjust its speed in the same way after a space displacement of $s_0$ and an adjusting time of $\tau$ to reach the preferred spacing for the new speed $v'$. NCM assumes that the continuous trajectory can be approximated by piecewise linear curves (the solid line segments in Figure 1 (a)), where the spacing $d_n(t)$ and speed $v_n(t)$ relationship of NCM is given by

$$d_n(t) = v_n(t)\tau + \delta \tag{1}$$

where $d_n(t) = x_{n-1}(t) - x_n(t)$, $\delta = s_0 + L_{veh}$, and $L_{veh}$ is the vehicle length, see Figure 1 (b). This linear relationship gives the solution of the kinematic-wave theory of Lighthill and Whitham (1955), and Richards (1956) with the triangular flow-density diagram, see Figure 1 (c). Therefore, under the car-following context, NCM is described by

$$v_n(t+\tau) = \min\left(v_{\max}, \frac{d_n(t)-\delta}{\tau}\right) \tag{2}$$

$$x_n(t+\tau) = x_n + v_n(t+\tau)\tau \tag{3}$$

where $v_{\max}$ is the maximum speed. NCM adopts a single wave speed $-\delta/\tau$ independent of traffic states. Traffic information, such as speeds, flows, spacing, etc., propagate unchanged along characteristic travelling at either $v_{\max}$ or $-\delta/\tau$ (Laval and Leclercq, 2010), which can be seen in Figure 1 (c).

Unfortunately, Newell model is not able to depict growth characteristics of traffic oscillations because the model is always stable and disturbance does not grow in this model. Thus, Laval and Leclercq (2010) have introduced the term $\eta_n(t)$ to describe non-equilibrium driving behavior. The model of Laval and Leclercq (LL model) is given as follows:

$$x_n^{free}(t) = x_n(t-\tau) + \min\left(v_{\max}, v_n(t-\tau) + a_m\left(1-v_n(t-\tau)/v_{\max}\right)\tau\right)\tau \tag{4}$$

$$x_n^{cong}(t) = x_n(t-\tau) + \max\left(d_n(t-\tau) + (1-\eta_n(t))v_{n-1}(t+\tau)\tau - \delta, 0\right) \tag{5}$$

$$x_n(t) = \min\left(x_n^{cong}(t), x_n^{free}(t)\right) \tag{6}$$

where $a_m$ is the maximum acceleration. LL model assumes that drivers in equilibrium will switch to non-equilibrium mode as soon as the leader $n$+1 decelerates, which is described by the dynamics of $\eta_n(t)$:

Before non-equilibrium:

$$\eta_n(t) = 1 \tag{7}$$

During non-equilibrium:

$$\eta_n(t) = \begin{cases} \eta + \varepsilon\tau, \eta_n(t) \leq \eta_T \text{ and after the non-equilibrium starts} \\ \eta - \varepsilon\tau, 1 \leq \eta_n(t) \leq \eta_T \text{ and after } \eta_T \text{ achieves} \end{cases} \tag{8}$$

After non-equilibrium:

$$\eta_n(t) = 1 \tag{9}$$

where $\varepsilon>0$, $\varepsilon=0$ and $\varepsilon<0$ give rise to three driver classes: timid, Newell, and aggressive drivers, respectively. Chen et al. (2012a, b) extended LL model based on empirical observations by considering driver category and reaction pattern. However, there is no essential distinction between the simulation results of the LL model and the model of Chen et al. with respect to the oscillation growth pattern. Figure 2 shows the simulation results of the LL model. It can be seen that the standard deviation of speed increases in a convex way in the front of the platoon, which obviously contradicts with the experimental results.

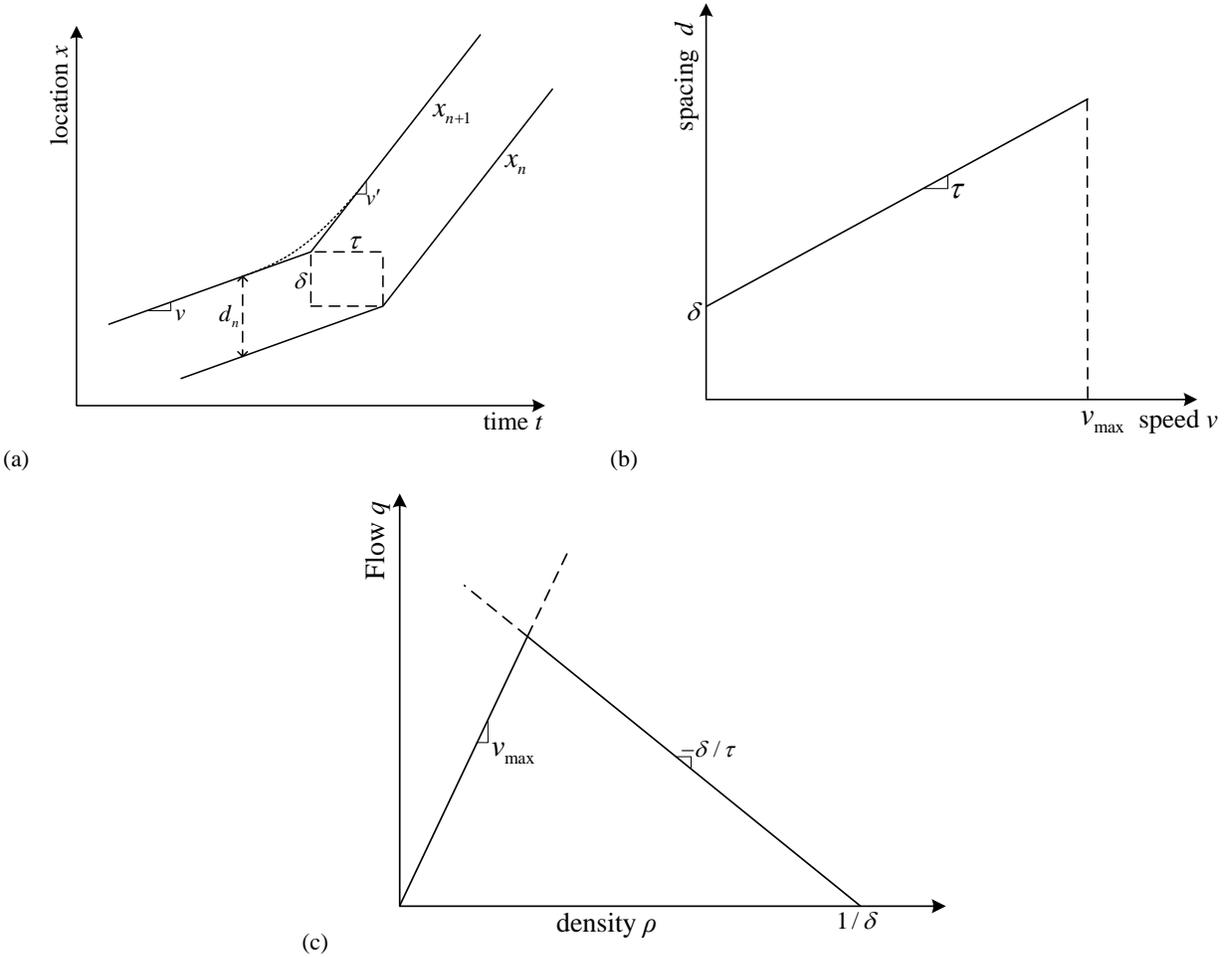

**Figure 1.** (a) Piecewise linear approximation to vehicle trajectories in NCM, (b) speed-spacing relationship for an individual vehicle in NCM, (c) density-flow relationship in NCM.

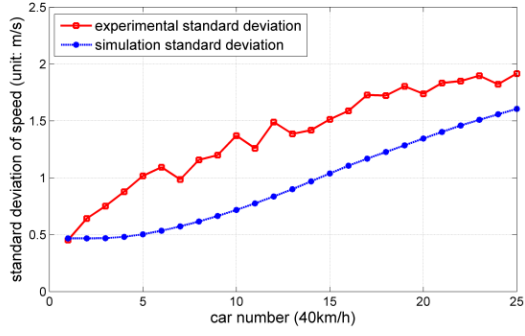
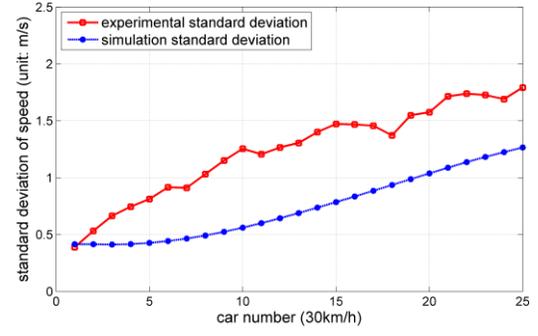

(a) (b)

**Figure 2.** Comparison of the experiment results (red lines) and simulation results of LL model (blue lines) with respect to the standard deviation of the speed of the cars. The car number 1 is the leading car. In the simulation, the values of parameters of LL model are set as: $\eta_T$=1.05, $\varepsilon$=0.01, $s_0$=1.5m, $L_{veh}$=5m, $v_{max}$ = 30m/s, $\tau$=1s, $a_m$ = 1.5m/s$^2$. Details of the simulation are given in section 6.1.

Recently, Laval et al. (2014) have proposed a stochastic desired acceleration model (SDAM), which incorporated the stochastic nature of driver's acceleration process into Newell model by adding a white noise to drivers' desired acceleration.

$$x_n^{free}(t) = x_n(t-\tau) + \max\left(\min\left(v_{max}\tau, \xi\right), 0\right) \quad (10)$$

$$x_n^{cong}(t) = x_{n+1}(t-\tau) - \delta \quad (11)$$

$$x_n(t) = \min\left(x_n^{cong}(t), x_n^{free}(t)\right) \quad (12)$$

where $\xi$ is generated by the Normal distribution with the mean $E$ and variance $V^2$ as follows:

$$E = v_{max}\tau - \left(1 - e^{-\beta\tau}\right)\left(v_{max} - v_n(t-\tau)\right)/\beta \quad (13)$$

$$V^2 = \frac{\sigma^2}{2\beta^3}\left(e^{-\beta\tau}\left(4 - e^{-\beta\tau}\right) + 2\beta\tau - 3\right) \quad (14)$$

where $\beta$ is the inverse of the relaxation time and $\sigma^2$ is the diffusion coefficient.

The parameter values of SDAM is calibrated to fit the concave growth pattern, see Table 1. Figure 3 compares the standard deviation of the speed of the cars between the simulated results of the SDAM and the experimental results. The model can simulate the concave growth of traffic oscillations pretty well. Now we examine whether the model can reproduce synchronized flow well. Figure 4(a) shows the flow-density diagram in which there is only one congested flow branch. Figure 4(b) shows a typical spatiotemporal diagrams of traffic flow on a circular road. One can see that the synchronized traffic flow cannot be simulated.

**Table 1.** Parameter values of SDAM.

| **Parameters** | $v_{max}$ | $\tau$ | $\beta$ | $\sigma$ | $s_0$ | $L_{veh}$ |
|---|---|---|---|---|---|---|
| **Units** | m/s | s | s$^{-1}$ | m/s$^2$ | m | m |
| **Value** | 30 | 1 | 0.03 | 0.6 | 1.5 | 5 |

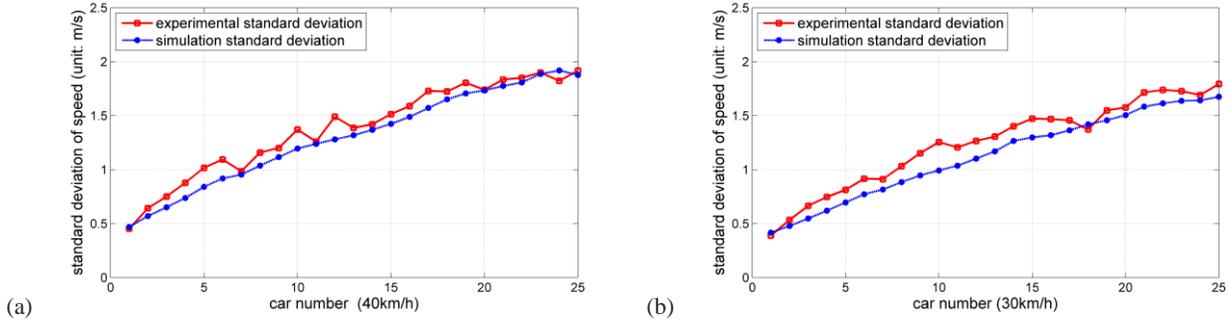

**Figure 3.** Comparison of the simulation results of SDAM (blue lines) and experiment results (red lines) with respect to the standard deviation of the speed of the cars. In the simulation, the values of parameters of SDAM model is set as Table 1. Details of the simulation is given in section 6.

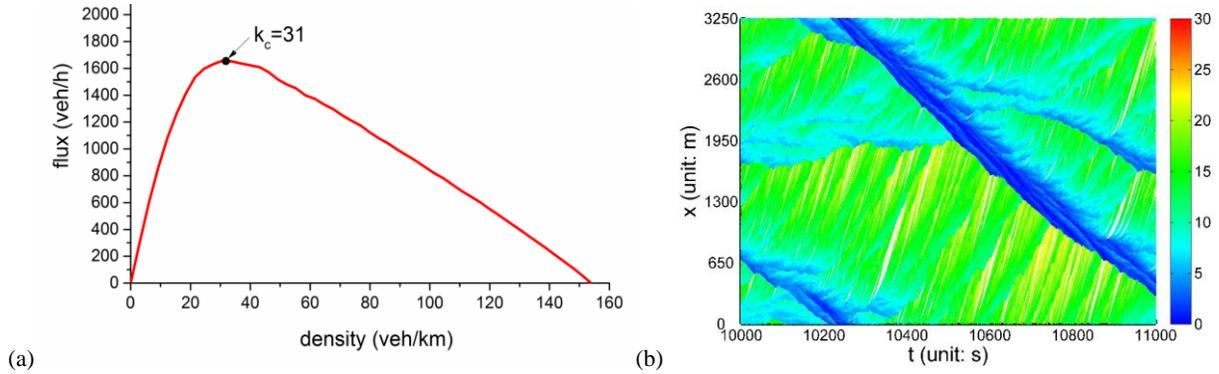

**Figure 4.** (a) Flow-density diagram of SDAM simulations on a ring road describing the averaged flow of traffic flow as a function of the global density (number of vehicles divided by the road length). The black points, marked by "$k_c = 31$" is the density with the maximum flux on the curves. (b) Spatiotemporal patterns of the SDAM on a circular road. The global density $k$=49 veh/km, $\sigma$=0.6.

## 3. Stochastic Newell car following model

Simulation results in section 2 indicate that: i) improving Newell model in a deterministic way cannot describe the oscillation growth pattern correctly; ii) improving Newell model in a stochastic way is potentially able to reproduce the oscillation growth pattern, but cannot reproduce the synchronized traffic flow. It seems that introducing stochastic factors with fixed strength overestimates fluctuations in synchronized flow and thus makes synchronized flow unstable. Based on the facts, we propose a stochastic Newell model considering the speed dependent stochasticity.

(i) Driver's acceleration is bounded, i.e. the velocity in the free flow calculated by

$$v_n^{free}(t+\tau) = \min\left(v_n(t) + a\tau, v_{\max}\right) \tag{15}$$

where $a$ is the average acceleration in the free driving style.

(ii) Driver's deceleration consists of the deterministic and stochastic components. The deterministic component is to ensure safety, i.e. the speed should be smaller than permitted by the spacing to avoid potential accident. The stochastic component is depicted by a randomization deceleration. For simplicity, the randomization deceleration is assumed to be the same as $a$. More importantly, the randomization probability $p_n$, is assumed to increase with the speed, i.e.

$$v_n(t+\tau) = \max\left(\min\left(v_n^{free}(t+\tau), \frac{d_n(t)-\delta}{\tau}\right) - a\mathrm{H}(r, p_n)\tau, 0\right) \tag{16}$$

$$p_n = p_a \frac{v_n(t)}{v_{\max}} \tag{17}$$

where H($r$, $p_n$) equals 1, if $r < p_n$; equals 0, otherwise. $r$ is a random number between 0 and 1.

(iii) When the speed is smaller than $v_{\text{jam}}$ (for simplicity, it is assumed $v_{\text{jam}} = a\tau$), the vehicle is supposed in the jam state. In such state, the randomization probability $p_n$ is presumed to be a constant $p_b$, which is applied to generate the realistic wide moving jams, i.e.

$$p_n = p_b \tag{18}$$

Since the randomization probability $p$ is related with driver's delay time via $\tau_{\text{del}} = \tau/(1-p)$, Equation (18) is applied to simulate the start delay when vehicles leave the jam. Moreover, as is well known, the propagation speed of wide moving jams falls into the interval [-20, -10]km/h. Therefore, adjusting $p_b$ can reproduce this different characteristic speed accordingly.

Based on these assumptions, the stochastic Newell car following model (SNCM) can be summarized as follows:

$$\tilde{v}_n(t+\tau) = \min\left(v_n(t) + a\tau, v_{\max}, \frac{d_n(t) - \delta}{\tau}\right) \tag{19}$$

$$v_n(t+\tau) = \max\left(\tilde{v}_n - a\mathrm{H}(r, p_n)\tau, 0\right) \tag{20}$$

$$x_n(t+\tau) = x_n(t) + v_n(t+\tau)\tau \tag{21}$$

where the randomization probability $p_n$ is defined as

$$p_n = \begin{cases} p_b & \text{if } v_n(t) < a\tau, \\ p_a \dfrac{v_n(t)}{v_{\max}} & \text{otherwise.} \end{cases} \tag{22}$$

## 4. Synchronized traffic flow simulation

This section investigates traffic flow on a circular road to identify whether SNCM can simulate the synchronized traffic flow discovered by Kerner (1998). The road length $L_{\text{road}} = 3250$m, which can contain at most 500 vehicles. Parameter values are shown in Table 2.

**Table 2.** Parameter values of SNCM.

| Parameters | $v_{\max}$ | $a$ | $\tau$ | $P_a$ | $P_b$ | $s_0$ | $L_{\text{veh}}$ |
|---|---|---|---|---|---|---|---|
| Units | m/s | m/s$^2$ | s | --- | --- | m | m |
| Value | 30 | 0.5 | 1 | 0.1 | 0.27 | 1.5 | 5 |

Two initial configurations are used in the simulations: 1) all vehicles are homogeneously distributed on the road; 2) all vehicles are distributed in a mega-jam. Figure 5 shows the resulting flow-density and speed-density diagrams, where the upper branch of $k_1 < k < k_2$ is from the initial homogeneous distribution, while the lower branch is from the initial mega-jam ($k_1 < k < k_2$). If the density is smaller than $k_1$, there is only free flow on the road since the speed is almost the maximum speed. In the upper branch $k_1 < k < k_2$, synchronized traffic flow begins to develop from free flow (Figure 6 (b)); increasing the density further, the region of synchronized flow will gradually increase until there is only synchronized traffic flow on the road (Figure 6 (c)). When the density is greater than $k_3$, the synchronized flow is unstable, and wide moving jams will appear (Figure 6 (d)). Note that the synchronized flow is spatiotemporally inhomogeneous, as observed in empirical data. For the lower branch formed from the initial mega-jam ($k_1 < k$), traffic will evolve to the state that wide moving jams and free flow coexist.

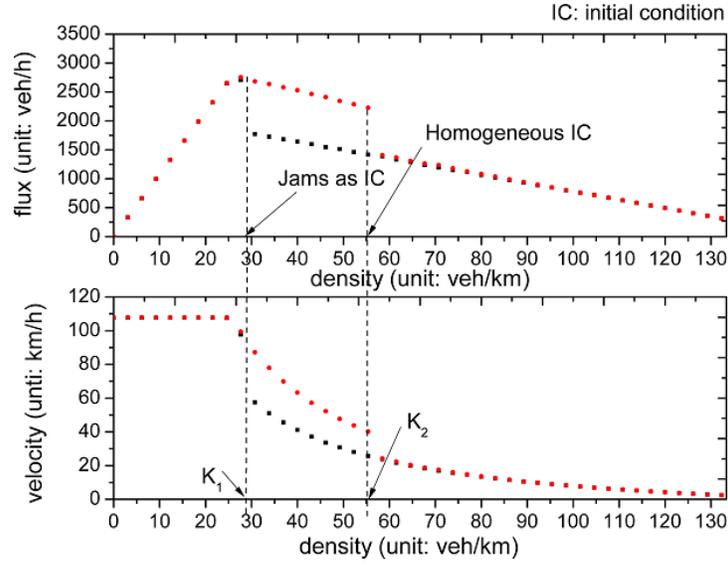

**Figure 5.** Space-averaged flow-density and speed-density diagrams of SNCM simulations on a ring road describing the average flow (or speed) of traffic flow as a function of the global density. "Homogeneous IC" represents the initially homogenous distribution of traffic, and "Jams as IC" represents the initially mega-jam distribution of traffic.

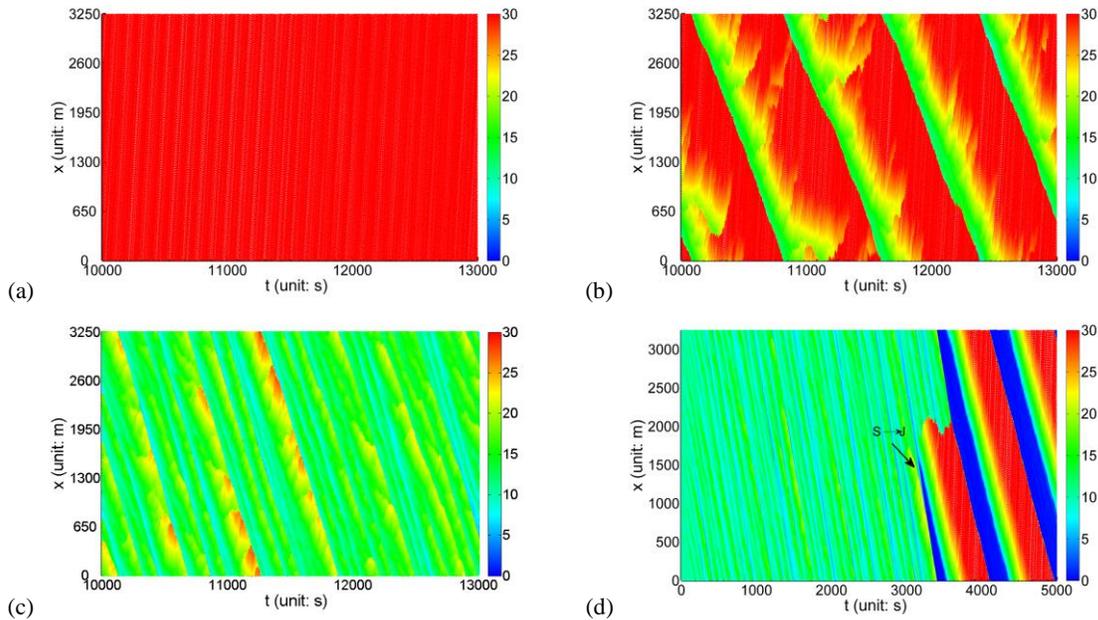

**Figure 6.** Spatiotemporal patterns of the SNCM on a circular road under different densities. In (a-d), the density $k$=25, 31, 52 and 62veh/km, respectively. In (a-d), the traffic starts from a homogeneous initial distribution. 'S→J' represents the transition from synchronized flow to jams. The color bar indicates speed.

## 5. Comparisons between simulation and NGSIM data

In this section, we simulate the empirical spatiotemporal traffic flow patterns in NGSIM data collected on a 640m-segment on southbound US 101 in Los Angeles, CA, on June 15th, 2005 between 7:50 a.m. and 8:35 a.m., see Figure 7 (a) and Figure 8 (a). Figure 7 (a) shows the emergences of jams, which arise with a regular period of about 2min, while Figure 8 (a) shows synchronized traffic flow upstream of the bottleneck.

Chen et al. (2012b) discovered that the rubbernecking caused by the clean-up work on Lane 1 between 7:50 a.m. and 8:05 a.m. is the most likely cause for traffic jams in this segment. The rubbernecking zone is located at [320, 420]m, where the road length $L_{road}$ = 640m. To simulate the rubbernecking, the method of Chen et al. (2012b) is applied. When vehicles enter this zone, at each simulation time step, they have a probability $r$ to rubberneck which will cause their speeds $v$ to decrease instantaneously to $v(1-p)$. Rubbernecking only can occur at most once for each vehicle in this zone.

Initially, the road section is assumed to be filled with vehicles uniformly distributed with density $\rho$ and their velocities are set to $v_0$. For the leading vehicle, it will be removed when it goes beyond the end of the road. The second car becomes new leading car and it moves freely. At the road section entrance, no vehicle is inserted. In this segment, the maximum speed is around 80km/h. Thus, $v_{max}$ of the SNCM is set as 80km/h. Other parameters are the same as Table 2.

Figure 7 (b) and Figure 8 (b) are the simulation results of SNCM, which show an attractive resemblance with Figure 7(a) and Figure 8(a). From Figure 7 (b), one also can see that not every rubbernecker will lead to a traffic jam. Furthermore, Figure 9 shows the relationship between the period of traffic jams with the rubbernecking parameters $r$ and $p$. Here period of traffic jams is calculated by

$$P = \frac{1}{L}\sum_{i=1}^{L}\frac{T}{l_i}. \qquad (23)$$

Here, $l_i$ denotes number of traffic jams detected by the virtual detector at location $x$=50m in simulation realization $i$. $T$ is simulation time of each realization, $L$ is number of simulation realization. In the simulation $T$ = 3000s and $L$ = 50. Notice that both $r$ and $p$ are negatively correlated with the period. The relation between the period and $r$ seems to be convex decreasing and approaches the stable values. The relation between the period and $p$ is also convex decreasing. The period also approaches the stable values. Figure 10 shows the relation between the period and the model parameter $p_b$, which reflects driver's start delay: a larger value means a higher delay. It seems that the period also increases with $p_b$ in a convex way.

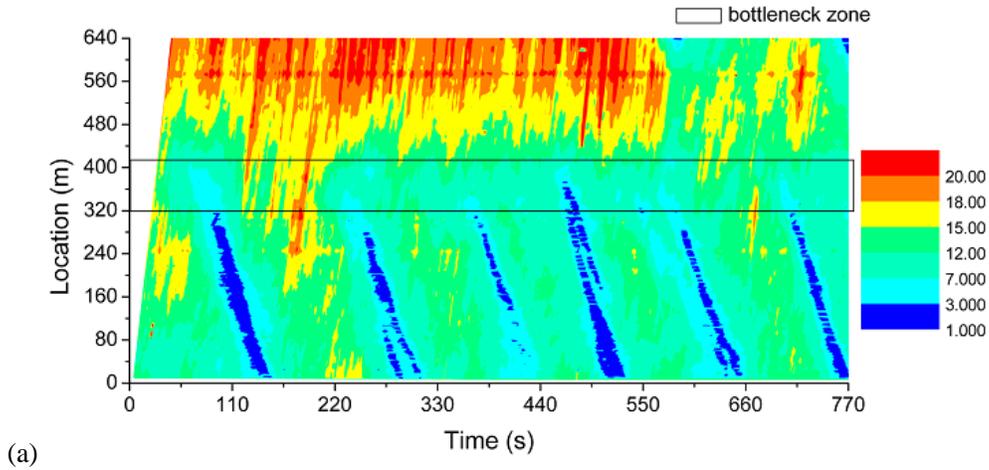

(a)

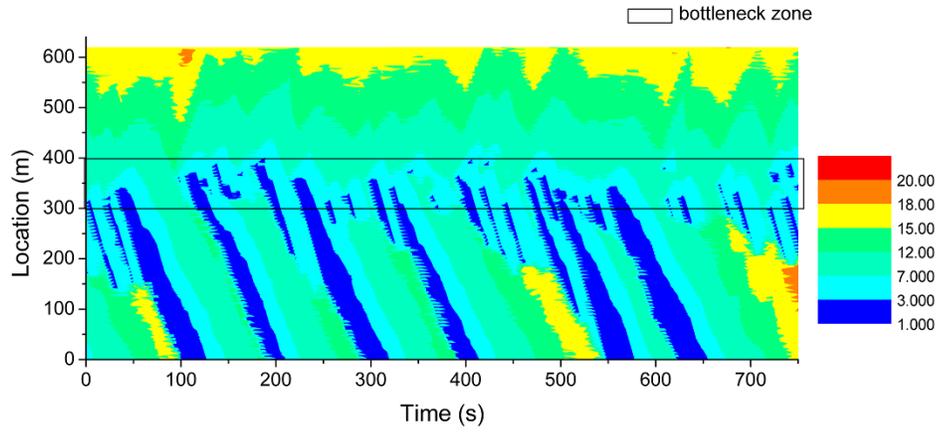

(b)

**Figure 7.** (a) Spatiotemporal speed patterns of the and Lane 1 from the NGSIM trajectory dataset collected on a 640*m*-segment on southbound US 101. (b) Simulation result of SNCM. *p*=0.8*, r*=0.012, $\rho$ = 61veh/km, $v_0$ = 30km/h. The color bar indicates speed.

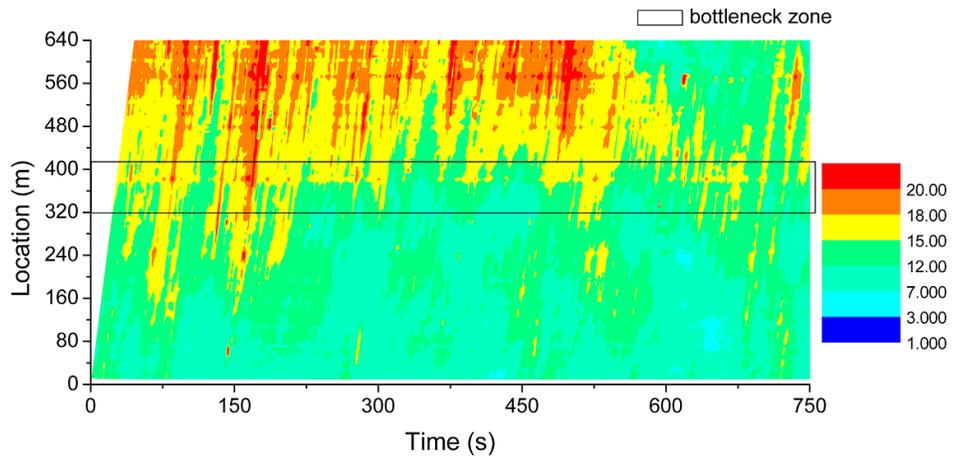

(a)

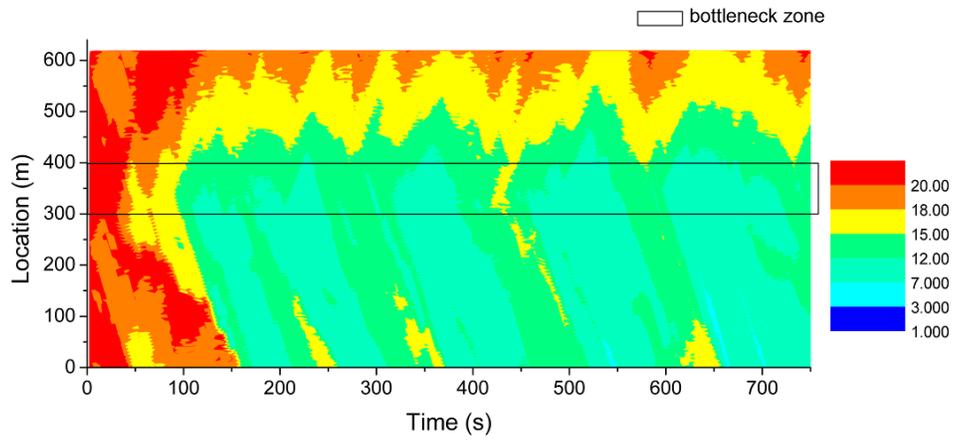

(b)

**Figure 8.** (a) Spatiotemporal speed patterns of the and Lane 4 from the NGSIM trajectory dataset collected on a 640m-segment on southbound US 101. (b) Simulation result of SNCM. *p*=0.12*, r*=0.03, $\rho$ = 37veh/km, $v_0$ = 80km/h. The color bar indicates speed.

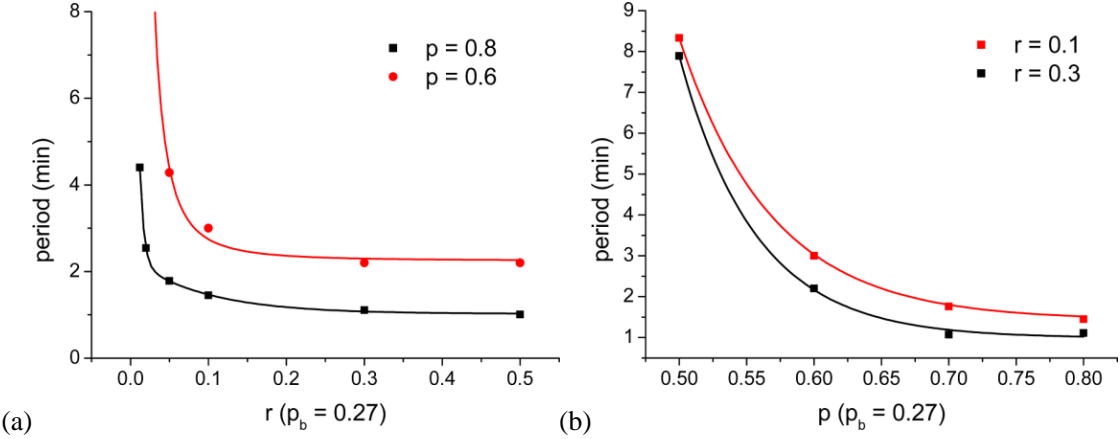

**Figure 9.** Traffic jams period versus rubbernecking parameters. The curves are the fitted lines and the symbols are the simulation results.

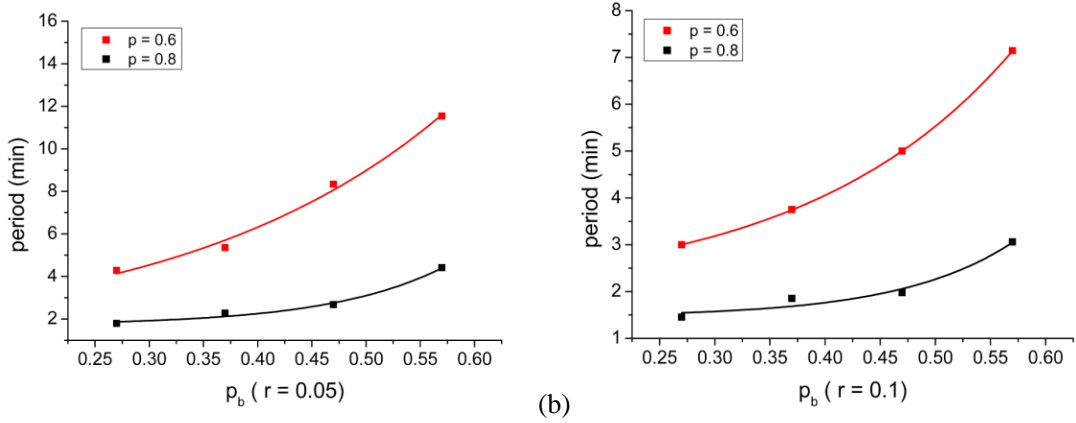

**Figure 10.** Traffic jams period versus model parameter. The curves are the fitted lines and the symbols are the simulation results.

## 6. Model calibration and validation

*6.1 Calibration and validation by the experimental data*

In this section, we calibrate and validate the SNCM by the car-following experimental data of Jiang et al. (2014) to examine whether the concave growth pattern of traffic oscillations can be quantitatively simulated or not. In the experiments, the driver of leading vehicle of the platoon was asked to move with different constant speed $v_{leading}$. At each $v_{leading}$, several runs were carried out. The stationary state data have been used to calculate the speed standard deviation, and the results have been averaged over the runs. As shown in Figure 10, $\sigma_{v,n}$ increases in a concave or linear way along the vehicles $n$ of the platoon.

We use the standard deviations $\sigma_{v,n}$ under $v_{leading}$= 50, 30 and 7km/h to calibrate the SNCM and $\sigma_{v,n}$ under $v_{leading}$= 40 and 15km/h for validation. The *RMSE* of the relative difference of the standard deviations between data and simulation is applied to measure the performance:

$$RMSE = \sqrt{\frac{1}{24}\sum_{n=2}^{25}\left(\frac{\sigma_{v,n}^{simu} - \sigma_{v,n}^{experi}}{\sigma_{v,n}^{experi}}\right)^2} \qquad (24)$$

where $n$ is the vehicle number, $\sigma_{v,n}^{simu}$ ($\sigma_{v,n}^{experi}$) represents the standard deviation of vehicle $n$ in the stationary state in the simulation (experiments). During the calibration, the objective is to minimize the average *RMSE* of the three datasets with $v_{leading}$= 50, 30 and 7km/h. The calibrated parameters are shown in Table 2.

**Table 3.** Calibrated parameter values of SNCM.

| Parameters | $v_{max}$ | $a$ | $\tau$ | $P_a$ | $P_b$ | $s_0$ | $L_{veh}$ |
|---|---|---|---|---|---|---|---|
| Units | m/s | m/s$^2$ | s | --- | --- | m | m |
| Value | 28.19 | 0.57 | 1.0 | 0.76 | 0.08 | 4.24 | 5 |

**Table 4.** The calibration and validation *RMSE* of the car following experiments.

| Model | | Calibration error | | | Validation error | |
|---|---|---|---|---|---|---|
| | $v_{leading}$ (km/h) | 50 | 30 | 7 | 40 | 15 |
| SNCM | *RMSE* | 0.19 | 0.07 | 0.18 | 0.06 | 0.22 |
| | Average *RMSE* | | 0.15 | | | 0.14 | |

The calibrated and validation errors are given in Table 3. It can be seen that: (i) all errors are in the reasonable range; (2) the validation error is at the same level as that of the calibration error, which indicates the good capability of the SNCM in predicting the observed growth pattern of the disturbances along the platoon, see Figure 11. Figure 12 shows that the formation and evolution of the simulated oscillations (right panels) look very similar to that of Jiang's experimental data (left panels).

Furthermore, to investigate the oscillation frequencies in the platoon and compare the simulation results with the experimental ones, we make a Fast Fourier transform analysis (FFT, Li, et al. 2010) of the time series of the velocity of the last vehicle in each platoon. As shown in Figure 13, one can see that the simulation results are roughly in agreement with the experimental ones. Specifically, the frequencies corresponding to the main peaks are roughly in agreement.

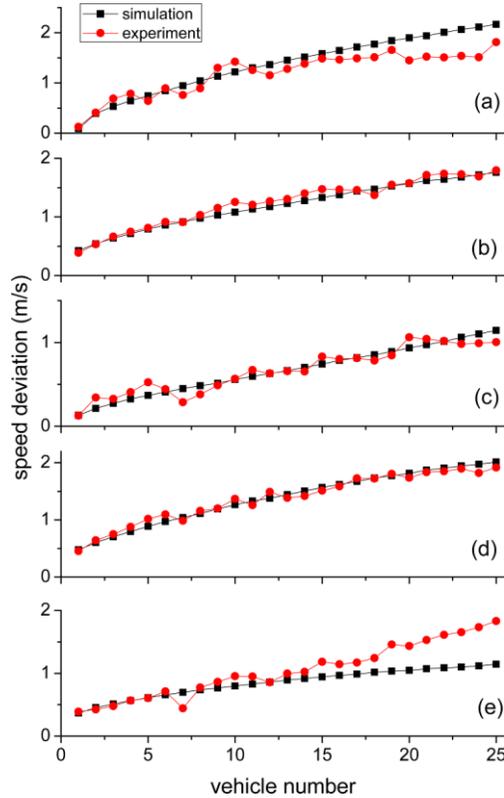

**Figure 11.** Comparison of the simulation results and experiment results of the standard deviation of the speed of the cars. The car number 1 is the leading car. In (a) - (e), the leading vehicle of the platoon is required to move with $v_{leading}$ =50, 30, 7, 40, 15km/h, respectively.

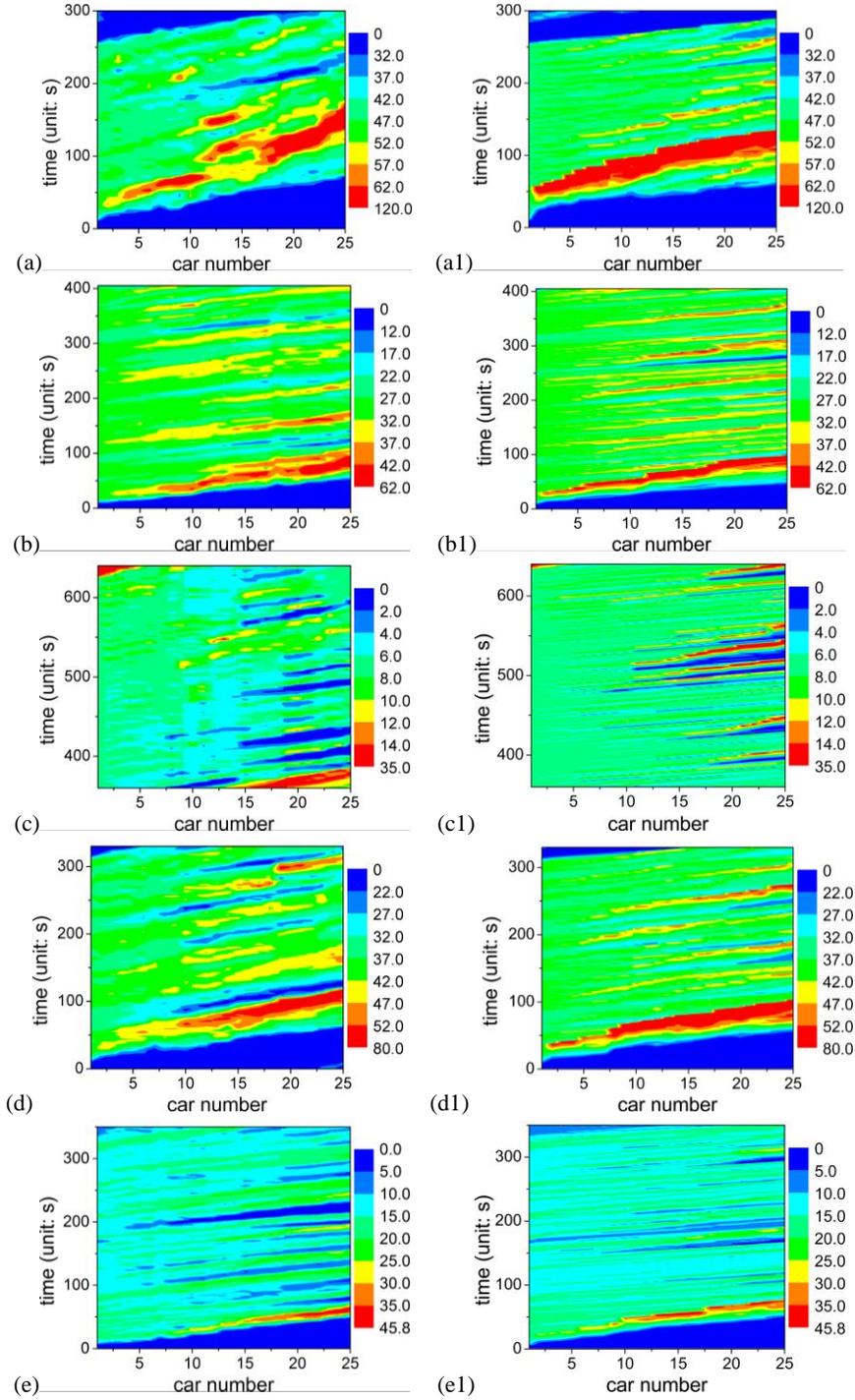

**Figure 12.** The spatiotemporal patterns of the platoon traffic. The car speed is shown with different colors (unit: km/h) as function of time and car number. The left and right Panels show the experimental results and the simulation results of SNCM. In (a, a1) - (e, e1), the leading vehicle of the platoon is required to move with $v_{leading}$ =50, 30, 7, 40, 15km/h, respectively. The color bar indicates speed.

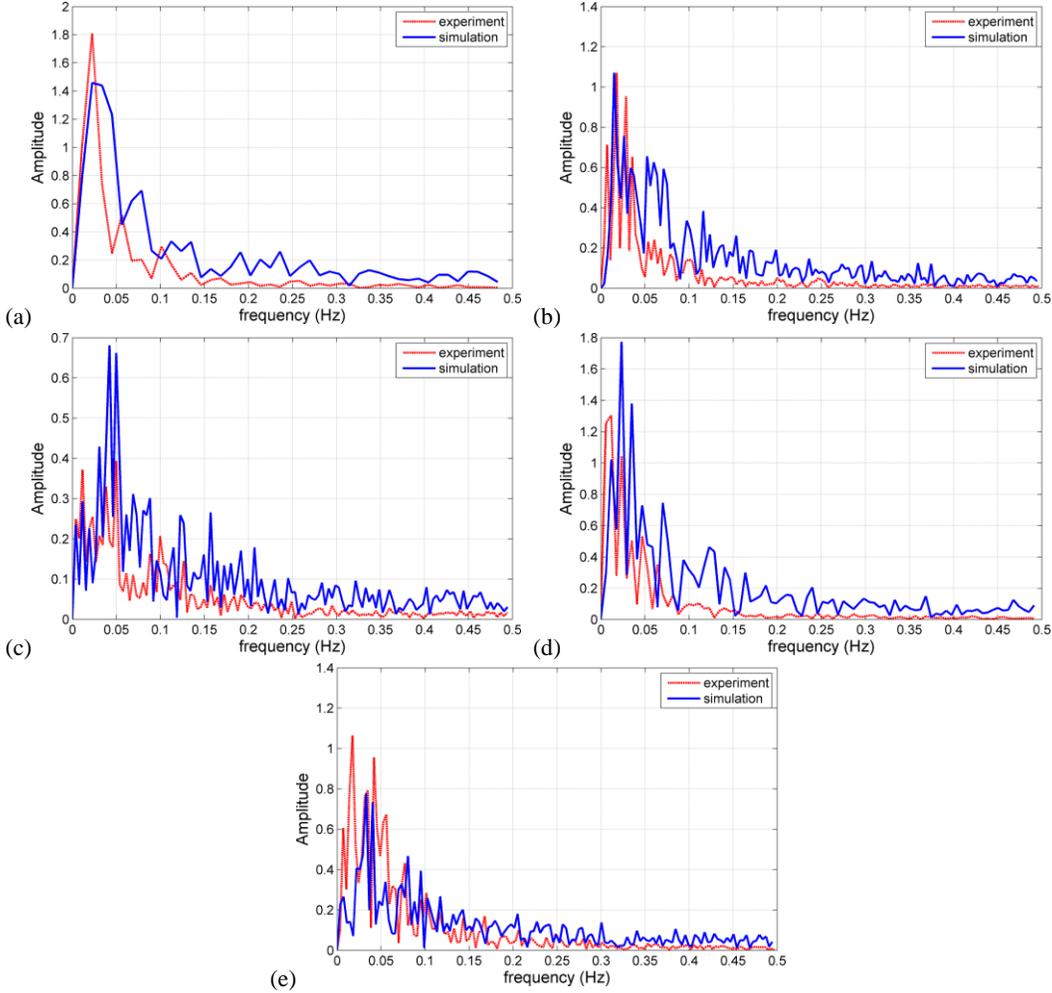

**Figure 13.** FFT spectra for detrended speed data of the last vehicle in the platoon. (a) - (e) corresponds to $v_{leading}$ = 50, 30, 7, 40, 15km/h.

*6.2 Calibration by NGSIM trajectory data*

We further calibrate and validate SNCM by the reconstructed NGSIM I80 dataset (Montanino and Punzo, 2015) to examine whether the car following behaviors of single vehicle can be quantitatively simulated or not. This dataset is from the northbound traffic on I80 in Emeryville, California, recorded from 4:00 p.m. to 4:15 p.m. on April 13, 2005. The reconstructed dataset is based on the raw dataset and has improved accuracy. Vehicle trajectories were recorded every 1/10 s. The area covered in these data sets has a length of about 500 m, includes six lanes and an on-ramp but does not include the off-ramp, see Figure 14. We use the data on the second lane, which includes both free flow and jams, and is therefore enough to test the performance of SNCM, see Figure 15. The following trajectory selection criteria are used to filter suitable trajectories from the empirical data:

1. The vehicle's leading car could not change lanes during the whole period.
2. There are at least 400 data points for the selected vehicle, i.e., a trajectory duration of 40 s or more.

After applying these criteria, 227 out of 404 vehicle trajectories were included in the calibration, see Figure 15. It can be seen that trajectories in free flow and jams are included. Therefore, these data are enough to evaluate whether the SNCM can describe the movement of the single vehicle well.

In the simulation setup, the calibration is performed by considering various leader-follower pairs and comparing the driving behaviors of the real and simulated follower obtained from the SNCM. The simulated relative speed and spacing are initialized by the real initial relative speed and gap. Then SNCM is applied to calculate the acceleration and thus the trajectory of the following vehicle. The spacing to the leading vehicle is computed as the difference between the simulated trajectory and the real position of the rear-bumper of the leading vehicle.

During the calibration process, the difference between the real driving behavior and the driving behavior obtained by the SNCM is quantified by the root-mean-square error (*RMSE*) of spacing, since optimizing with respect to spacing automatically reduces the average relative speed errors.

$$RMSE = \sqrt{\frac{1}{N}\sum_i (\frac{d_i^{\text{sim}} - d_i^{\text{emp}}}{d_i^{\text{emp}}})^2} \tag{25}$$

where $d_i^{\text{sim}}$ is i$^{\text{th}}$ simulated value; $d_i^{\text{emp}}$ is corresponding observed value from trajectory data; $N$ is number of observations. Calibration aims to minimize *RMSE* by choosing an "optimal" set of parameter values for SNCM. To find the optimal parameter values, the Genetic Algorithm (GA) provided by Matlab is applied. Since SNCM is a stochastic model, for each parameter value set generated by GA, the simulation is carried out *M* runs. Then the averaged *RMSE* is calculated to evaluate the performance of this value set. We found *M* =100 is enough to obtain a stable result.

The mean *RMSE* is 0.19. Figure 16 shows the calibration results. It can be seen that 8%, 48% and 25% of *RMSE* fall into the range [0, 0.1], [0.1, 0.2] and [0.2, 0.3], respectively, which means 81% of *RMSE* is less than 0.3. For illustrations, one set of the simulated sample versus trajectory data are shown in Figure 17. We have performed the cross validation approach (Martin and Kesting, 2013) to assess the robustness of the calibrated parameter values, since the model parameters are driver specific. Cross validation is performed as follows. We calibrated the parameters of vehicle *n* using trajectories of vehicles *n* and *n*-1. Then we make validation using the trajectories of other 226 vehicles. For each calibrated parameter value set, we used it to simulate all 227 vehicle trajectories. Then, the *RMSE* of all trajectories are calculated. Apparently, it is impossible that all *RMSE* values fall into the reasonable error range (smaller than 0.2 or 0.3), since Hamdar et al. (2015) have indicated that considerable contribution of inter-driver heterogeneity exists in I-80 data. Therefore, we evaluate the validation results by the index $N_r$, meaning that there are $N_r$ values of *RMSE* belonging to the reasonable error range for each calibrated parameter value set, i.e. this value set can successfully simulate $N_r$ vehicle trajectories. Obviously, a large value of $N_r$ means good performance of SNCM. Figure 18 shows the result when the reasonable error range is defined as *RMSE*≤0.3, which demonstrates that 81%, 67% and 14% calibrated parameter value sets can reproduce at least 20, 40 and 70 trajectories, respectively. If more stringent standard is considered, i.e. reasonable error range is defined as *RMSE*≤0.2, it has been shown in Figure 19 that 84%, 67% and 44% calibrated parameter value sets can simulate at least 5, 10 and 15 trajectories, respectively.

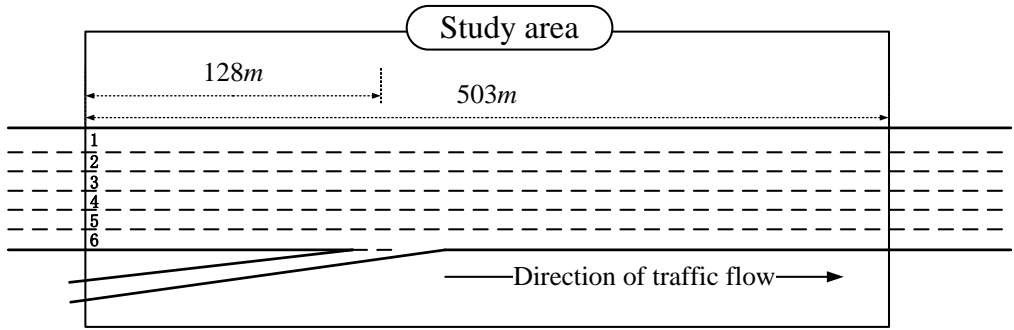

**Figure 14.** The sketch of I-80 study area.

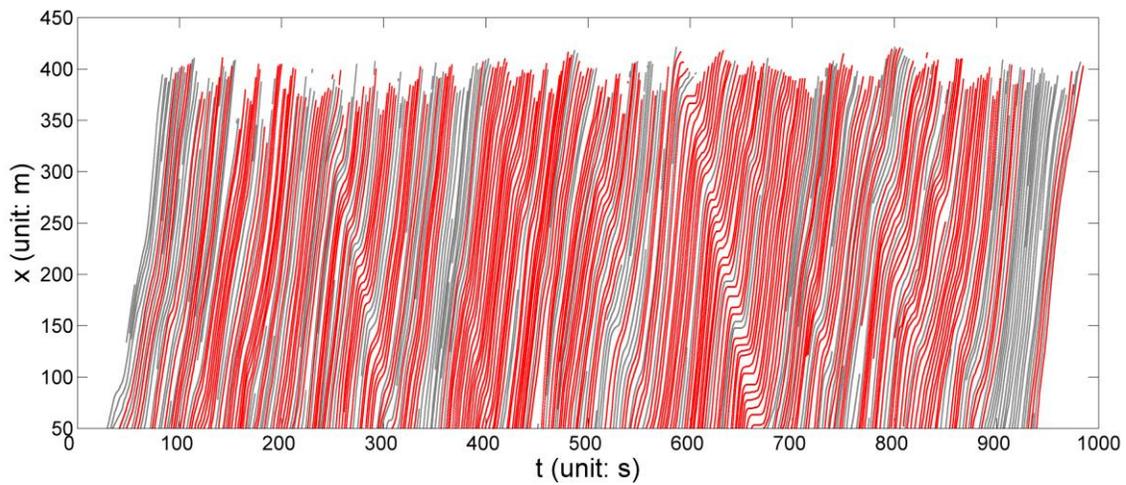

**Figure 15.** Time-space diagram of the second lane from the reconstructed I-80 dataset. The red trajectories are used in calibration.

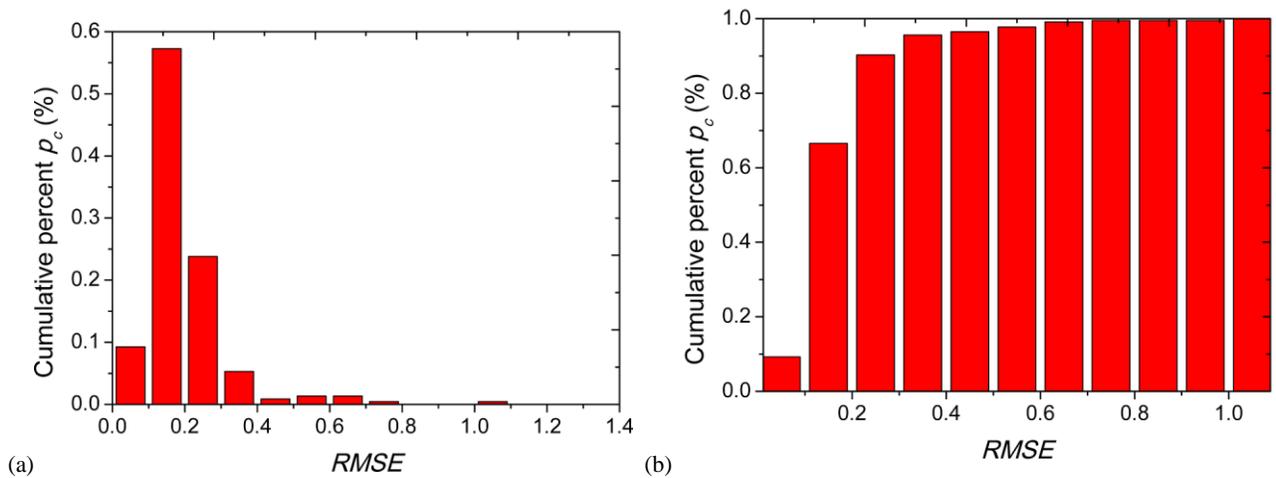

**Figure 16.** Calibration distribution across vehicles.

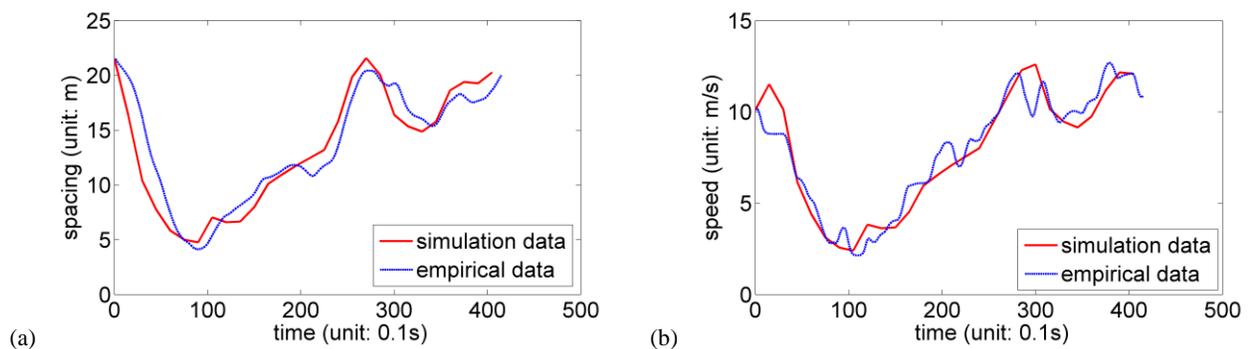

**Figure 17.** Simulated versus observed spacing and speed of Vehicle 1772. The simulation data are the 50 percent quantiles in 100 runs.

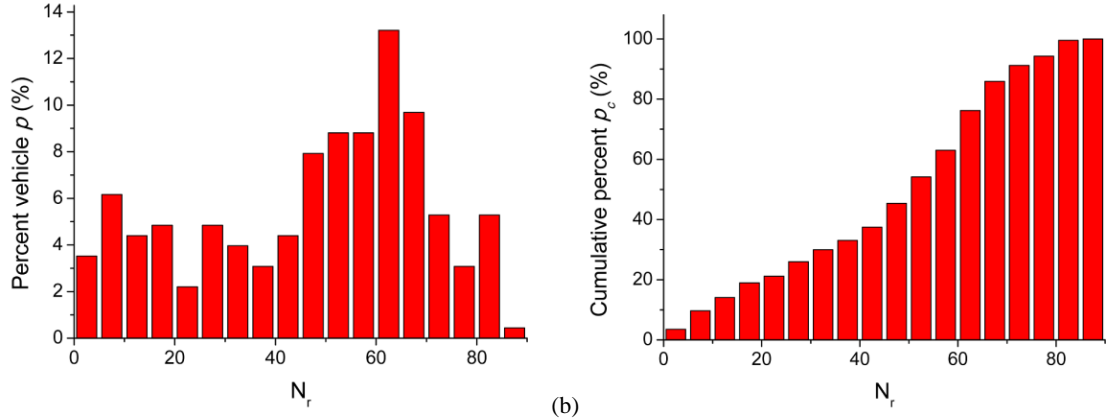

**Figure 18.** Validation distribution across vehicles for $RMSE \leq 0.3$. (a) there are $p$ percent of calibrated parameter value sets can simultaneously reproduce at least $N_r$ trajectories of the reconstructed I-80 data applied here with the $RMSE \leq 0.3$. (b) there are 1-$p_c$ percent of calibrated parameter sets can simultaneously reproduce more than $N_r$ trajectories of the reconstructed I-80 data applied here with the $RMSE \leq 0.3$.

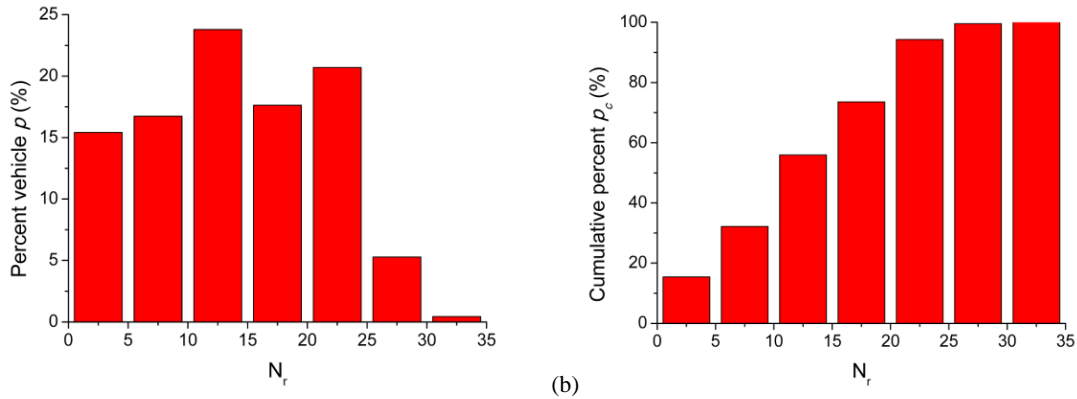

**Figure 19.** Validation distribution across vehicles for $RMSE \leq 0.2$. (a) there are $p$ percent of calibrated parameter value sets can simultaneously reproduce at least $N_r$ trajectories of the reconstructed I-80 data applied here with the $RMSE \leq 0.2$. (b) there are 1-$p_c$ percent of calibrated parameter sets can simultaneously reproduce more than $N_r$ trajectories of the reconstructed I-80 data applied here with the $RMSE \leq 0.2$.

## 7. Conclusion

Traffic flow exhibits complex spatiotemporal phenomena, among which, synchronized flow and concave growth pattern of oscillations are two important features. However, many well-known models fail to reproduce these two features well. This paper established an improved Newell car following model via introducing speed dependent stochasticity. Simulation results show that the empirical characteristics of the synchronized traffic flow and the periodic property of traffic oscillations observed from the NGSIM data can be generated. Calibration and validation with respect to experimental data shows that the concave growth pattern has been simulated successfully. Calibration and cross validation with respect to NGSIM vehicles' trajectories data also exhibit good performance of the model.

Traditional car-following studies usually ignore stochasticity. However, as revealed in recent studies (Kim and Zhang, 2008; Laval et al., 2014; Jiang et al., 2015; Tian et al., 2017; Sheu and Wu, 2015), the stochasticity plays an important role in traffic flow. This paper further shows that speed dependent stochasticity might be a nontrivial component in traffic flow. In our future work, more experimental and empirical traffic data need to be collected to reveal the effect of the speed dependent stochasticity on features such as traffic breakdown from free flow to synchronized flow.

## Acknowledgements:

This work is supported by National Key R&D Program of China (No. 2017YFC0803300). JFT was supported by the National Natural Science Foundation of China (Grant No. 71771168). RJ was supported by the National Natural Science Foundation of China (Grant Nos. 71621001, 71631002). Correspondence and requests for materials should be addressed to the RJ and BJ.